**Partial volume correction improves theranostic $^{124}$I/$^{131}$I-CLR1404 tumor dosimetry in xenograft models of head and neck cancer**

Ian R. Marsh[1], Chunrong Li[2], Joseph Grudzinski[1], Justin Jeffery[1], Colin Longhurst[3], David P. Adam[1], Reinier Hernandez[1], Jamey P. Weichert[4], Paul M. Harari[2], Bryan P. Bednarz[1]

[1]Department of Medical Physics, School of Medicine and Public Health, University of Wisconsin-Madison, Madison, WI 53705; [2]Department of Human Oncology, School of Medicine and Public Health, University of Wisconsin-Madison, Madison, WI 53792; [3]Department of Biostatistics and Medical Informatics, School of Medicine and Public Health, University of Wisconsin-Madison, Madison, WI, 53792; [4]Department of Radiology, School of Medicine and Public Health, University of Wisconsin-Madison, Madison, WI 53792;

**Corresponding Author:** Bryan P Bednarz, Department of Medical Physics, 1005 WIMR, 1111 Highland Avenue, Madison WI 53705; +1-608-262-5225; bbednarz2@wisc.edu

**First Author:** Ian R Marsh (graduate student), Department of Medical Physics, 1005 WIMR, 1111 Highland Avenue, Madison WI 53705; +1-608-698-0205; imarsh@wisc.edu

**Grant Support:** SPORE P50DE026787, Training Grant T32CA009206, U01 CA233102-01, P01 CA250972-01

**Word Count:** 7159

**Running title:** CLR 124/131 targets HNC

**ABSTRACT**

Combination radiopharmaceutical and external beam radiotherapy offers the potential to diminish locoregional toxicity that remains dose-limiting in the conventional treatment paradigm for recurrent head and neck cancer (HNC). In this study, we investigated the tumor targeting capacity of $^{131}$I-CLR1404 (CLR 131) in various HNC xenograft mouse models and the impact of partial volume correction on theranostic dosimetry based on $^{124}$I-CLR1404 (CLR 124) PET/CT imaging.

**Methods:** Mice bearing flank tumor xenograft models of HNC (6 murine cell line- and 6 human patient-derived) were intravenously administered 6.5-9.1 MBq of CLR 124 and imaged five times over the course of six days using microPET/CT. *In vivo* tumor uptake of CLR 124 was assessed and partial volume corrections (PVC) for $^{124}$I were applied using a novel preclinical phantom. Using subject-specific theranostic dosimetry estimations for CLR 131 based on CLR 124 imaging, a discrete radiation dose escalation study (2, 4, 6, and 8 Gy) was performed to evaluate tumor growth response to CLR 131 relative to a single fraction of XRT (6 Gy).

**Results:** PET imaging demonstrated consistent tumor selective uptake and retention of CLR 124 across all HNC xenograft models. Peak uptake of 4.4 ± 0.8% and 4.2 ± 0.4% was observed in SCC-22B and UW-13, respectively. PVC application increased uptake measures by 47-188% and reduced absolute differences between *in vivo* and *ex vivo* uptake measurements from 3.3 to 1.0 %IA/g. Tumor dosimetry averaged over all HNC models was 0.85 ± 0.27 Gy/MBq (1.58 ± 0.46 Gy/MBq with PVC). Therapeutic CLR 131 studies demonstrated a variable, but linear relationship between CLR 131 radiation dose and tumor growth delay ($p < 0.05$).

**Conclusion:** CLR 131 demonstrated tumoricidal capacity in preclinical HNC tumor models and the theranostic pairing of CLR 124/131 presents a promising new treatment approach for personalizing administration of CLR 131.

**Keywords:** radiopharmaceutical therapy, theranostics, dosimetry, CLR1404, head and neck cancer

**INTRODUCTION**

Approximately 50% of patients with head and neck cancer (HNC) manifest recurrence following initial treatment, with the majority of these recurrences being loco-regional (mouth, throat, neck)(*1–4*). Although some of these patients remain potentially curable with further local treatment approaches (surgery, radiation, chemoradiation), retreatment is technically challenging and accompanied by a significant risk of damage to normal tissues. Surgery is often limited by tumor adherence to critical structures (base of skull, neurovascular structures), whereas external beam radiation therapy (XRT) is often limited by normal tissue tolerance (spinal cord, bone, cartilage). Although frequently warranted in the attempt to provide disease control, HNC retreatment can induce profound adverse effects on patient health-related quality of life (QOL) (*5–10*). Therefore, improved treatment approaches for patients with loco-regional HNC recurrence are needed.

One promising approach to treat loco-regional HNC recurrence is to deliver a targeted radiopharmaceutical agent as an adjuvant to XRT thereby potentially limiting additional dose to surrounding normal tissues. This would allow for a decrease in the total dose applied via external beam radiation, thereby reducing side effects, while maintaining or potentially increasing tumor control. Such an approach is of particular significance in HNC because surgery and radiation often compromise normal salivary and swallow function with a powerful adverse impact on patient QOL (*11–15*). A promising agent that can be used in the adjuvant setting to treat HNC is CLR 131, which is an iodinated alkyl phosphocholine (APC) analog with broad cancer targeting abilities (*16*). APC analogs are internalized preferentially by cancer cells via lipid rafts which are specialized plasma membrane microdomains that spatially organize critical signaling molecules. Malignant cells possess lipid rafts at concentrations 6-10x higher than normal cells, which is considered a major reason for the almost universal tumor selectivity of these agents regardless of histopathological origin (*17,18*). The theranostic properties of CLR can be exploited for tumor imaging by labeling with $^{124}$I (CLR 124) as well as treatment by labeling with $^{131}$I (CLR 131). Particularly in combination with conventional modalities of cancer treatment, such as XRT, this strategy affords a unique approach to improve therapeutic outcome.

Preclinical studies have demonstrated selective uptake and retention of CLR compared with normal or hyperplastic tissues or benign tumors in over 50 human tumor xenografts and spontaneous animal tumor models including several HNCs (*16*). Despite these efforts, there remains a need to further characterize the tumor dosimetry of CLR in HNC models as an adjuvant therapy with XRT. Accurate dosimetry in mice is required to extrapolate the potential efficacy of this combined approach to humans. Our group has used PET/CT imaging of CLR124 and Monte Carlo radiation transport methods to characterize the biodistribution and estimate organ and tumor absorbed doses in mice from CLR 131, thus providing useful information in the early stages of the drug development and clinical translation (*19–21*). Corrections for partial volume effects (PVEs) in PET/CT images can also improve the accuracy of dosimetry studies. PVEs lead to an underestimate of tumor uptake in PET/CT images, particularly in small tumor lesions, which leads to an underestimate of tumor absorbed doses. We and others have shown that "hot sphere" phantoms can be used to produce recovery coefficient (RC) correction models leading to more accurate tumor dose estimates in preclinical dosimetry studies (*22,23*).

In this work, we evaluated the tumor-selective uptake of CLR 124 in murine and patient derived xenograft models of several HNCs. We introduced a novel 3D printed "hot spheres" phantom to produce a RC correction model on a cohort of mice bearing HNC xenografts. The RCs derived from the phantom were applied to CLR 124 PET/CT images of HNC xenografts and compared to biodistribution data from the same tumor-bearing mice. Monte Carlo dosimetry was performed on both corrected and uncorrected image datasets to demonstrate the dosimetric impact of accounting for PVEs.

**MATERIALS AND METHODS**

**Radiopharmaceuticals**

Methods for synthesis of CLR1404 and radiolabeling with $^{124}$I (CLR 124) or $^{131}$I (CLR 131) have been described previously (*16,24*). Unconjugated CLR1404 and CLR 131 were kindly provided by Cellectar Biosciences (Florham Park, NJ). $^{124}$I was obtained from IBA Molecular North America. CLR 124 was prepared via isotope-exchange reaction (*25*) by Dr. Jamey Weichert. All radiopharmaceuticals were prepared under good manufacturing practice guidelines.

**Human Head and Neck Cancer Models**

*In vivo* uptake of CLR 124 and theranostic CLR 131 dosimetry was studied in six human HNC cell lines (SCC-2, SCC-6, SCC-22B, SCC-47, SCC-1483, and Tu-138) and six HNC patient-derived xenografts (PDX) (UW-1, UW-13, UW-22, UW-25, UW-36, and UW-64). Cell lines SCC-1, SCC-6, and SCC-47 were obtained from MilliporeSigma (Burlington, MA) whereas SCC-2, SCC-1483, and Tu-138 were provided by Dr. Henning Bier via Dr. Thomas Carey (Technical University Munich), the Marnett Group (Vanderbilt University), and Dr. Jennifer Grandis (University of Pittsburgh), respectively. All HNC cell lines were cultured as previously described (*26*). The authenticity of each cell line was verified within 6 months of cell use via short-tandem repeat profiling (University of Wisconsin-Madison Pathology Core Lab).

PDX models can provide an advantage over established cancer cell lines in predicting patient response to targeted drugs because they more accurately reflect the biological characteristics of human cancers. Briefly, tumor tissue was collected at the time of surgery or staging biopsy from consenting patients (IRB#2016-0934) and transferred to the laboratory in ice-cold culture media containing DMEM. The sample was then minced into less than 1 mm$^3$ pieces under sterile conditions and mixed in a 1:1 ratio with matrigel and implanted into 4-6-week-old NOD-SCID gamma mice (NSG, Jackson Laboratories). Subsequent passages are made into either NSG or athymic nude mice in a similar fashion. PDX models used in this work were developed and characterized as previously described (*27,28*). Two of six PDX models (UW-1 and UW-36) have been validated as HPV+ via p16 staining, a status previously demonstrated to enhance radiation sensitivity (*29*).

**Mice and xenograft tumor models**

All animal studies were conducted under NIH guidelines and Institutional Animal Care and Use Committee approved protocols. Human tumor xenografts were established in 4-6-week-old female athymic nude mice (Envigo Bioproducts) on the right and/or left hind flank. For cell line derived tumors, a 200 µL cell suspension containing 1-2 million tumor cells was injected subcutaneously. For the PDX models, tumor tissue was harvested from NSG mice bearing the PDX of interest and implanted into athymic nude mice. Tumor growth was monitored twice per week via digital caliper measurements of the long (*L*) and short (*W*) axis of tumors and volume determined using the ellipsoid volume formula $\left(V = \frac{\pi}{6}LW^2\right)$. *In vivo* imaging studies were performed when tumors reached a volume of ~400 mm$^3$.

**PET Imaging and Biodistribution of CLR 124**

Mice bearing flank tumor xenografts (n= 2-6 per tumor model, n = 48 total) were administered 6.5 or 9.1 MBq of CLR 124 via lateral tail vein injection. Sequential CT and PET scans were acquired in an Inveon micro-PET/CT scanner (Siemens Preclinical Systems, Knoxville, TN) at 1, 24, 48, 72, and 144 h after injection. Prior to each scan, mice were anesthetized with isoflurane (4% induction, 2% maintenance) and placed in a prone position on the heated scanner bed. CT scans (80 kVp; 1000 mAs; 275 ms; 220 angles) were reconstructed via filtered back projection using the system software with a Shepp-Logan filter to a 0.2 x 0.2 x 0.2 mm$^3$ resolution. List mode PET scans consisting of 40 million coincidence events per mouse (energy window; 350-650 keV) were reconstructed using a two-dimensional ordered subset expectation maximization (OSEM2D) algorithm with 16 subsets and 4 iterations to a 0.8 x 0.8 x 0.8 mm$^3$ resolution. The resulting PET and CT image volumes were coregistered and corrections for attenuation, normalization, dead-time, and scatter were applied using the system software. Region of interest (ROI) analysis and quantification was performed with contours drawn on the anatomic CT image volumes. Quantitative *in vivo* data is expressed as percent injected activity per gram (%IA/g), assuming unit density of tissue. Following the final 144 h time point, a subset of mice was sacrificed, and tumor tissue was collected to corroborate the accuracy of *in vivo* measurements. Tumors were wet-weighed and activity was counted in a Wizard 2 automated γ-counter (Perking Elmer, MA) to calculate the %IA/g *ex vivo*.

**Partial Volume Correction Phantom**

We assessed the impact of PVE on $^{124}$I PET-based tumor uptake and dosimetry in this work with a 3D printed partial volume correction phantom (Phantech, Madison, WI). The phantom contains a single-channel series of 9 spherical voids ranging from 3.7 to 24.8 mm in diameter. Spheres in the phantom were filled with $^{124}$I solution at 298 kBq/ml, comparable to *in vivo* CLR 124 observations, and scanned with micro-PET/CT using the same imaging parameters used for mice. ROIs were drawn on the CT image volume to measure observed activity concentration (AC) and calculate recovery coefficients (RC) at each sphere size using Equation 1.

$$RC = \frac{AC_{measured}}{AC_{true}}$$  Eq. 1

The relationship between RC and sphere diameter was then fit to a sigmoidal function to allow for calculation of RC at any sphere diameter corresponding to measured tumor volume delineated in imaging studies.

Correcting for PVE in a defined tumor volume at a single point in time is straightforward. Propagating PVC through to tumor dosimetry, a product of the temporal integration of the spatially varying dose distribution, requires consideration of the entire time-activity curve. The area under the time activity curve of CLR 131 (AUC) incorporating PVC is given by Eq. 2, where RC is a function of the tumor volume over the course of the imaging series. The absorbed dose (AD) for a given tumor can then be corrected by the ratio of corrected and uncorrected AUC (Eq. 3). Assuming a constant tumor volume over the course of an imaging study, this correction factor is reduced to $RC(\bar{V})^{-1}$, where $\bar{V}$ is the mean tumor volume.

$$AUC_{PVC} = \int_0^\infty RC(V(t))^{-1} \cdot A_{CLR\,131}(t)\, dt \qquad \text{Eq. 2}$$

$$AD_{PVC} = AD \cdot \left(\frac{AUC_{PVC}}{AUC}\right) \qquad \text{Eq. 3}$$

### Theranostic Dosimetry in Tumors

Dosimetry for CLR 131 in this study was assessed via *in vivo* CLR 124 imaging using an in-house internal dosimetry platform as previously described (*19,20,30*). PET and CT image volumes at each time point were coregistered and resampled to an intermediate voxel resolution (0.4 x 0.4 x 0.4 mm$^3$) using a Mitchell-Netravali resampling filter. CLR 131 activity distributions were generated from CLR 124 PET images, correcting for differences in radioactive decay rates of $^{124}$I and $^{131}$I. Absorbed dose rate distributions for CLR 131 were generated at each time point using Geant4 (version 9.6) Monte Carlo simulations using the CT and PET volumes to represent the geometry and source distributions, respectively. Approximately 15 million $^{131}$I decays were simulated in mice at each time point (over 5,000 decays per source voxel) to achieve mean relative error in tumor ROIs of less than 2%. The mean absorbed dose rate in tumor ROIs was determined at each time point and the integral absorbed dose was calculated using a trapezoidal fit between imaging time points and assuming physical decay from the final time point. The assumption of prolonged retention of CLR 131 beyond the 144 h imaging time point is supported by observations from previous preclinical and clinical studies (*30–32*).

**Radiopharmaceutical therapy with CLR 131**

Tumor growth delay was assessed in SCC-2 and Tu-138 xenograft mouse models (n = 6-8) following the administration of CLR 131 RPT tailored to achieve tumor dose levels of 2, 4, 6, and 8 Gy. Theranostic dosimetry studies of CLR 124 in representative tumor bearing mice informed the prescription of CLR 131 to SCC-2 (0.78 Gy/MBq) and Tu-138 (1.16 Gy/MBq) tumors. Mice bearing SCC-2 xenografts received 2.6, 5.1, 7.7, or 10.2 MBq while mice bearing Tu-138 xenografts received 1.8, 3.6, 5.4, or 7.2 MBq of CLR 131. Mice receiving cold CLR1404 were used as control. An additional group of mice in each cohort were treated with 6 Gy of tumor-targeted XRT using an X-RAD 320 irradiator (Precision X-Ray Inc, North Branford, CT). RPT experiments were carried out once tumors achieved a volume of ~400 mm$^3$. Tumor volume was assessed via caliper measurements twice weekly for 4 weeks following treatment.

Clonogenic assays were performed for SCC-2 and Tu-138 cell lines to characterize in vitro response to radiation. Cells were seeded to 6-well plates at specified quantities (n=3) and irradiated by 0, 2, 4, 6, or 8 Gy using the X-RAD 320 irradiator. After 10-14 days incubation, cells were fixed and stained with 5% crystal violet in methanol and counted. Clonogenic survival was fit to a linear-quadratic model using non-linear least-squares to quantify characteristic radiobiological factors (*33*).

**Statistical Analysis**

To assess the reproducibility and accuracy of partial volume corrected (PVC) in-vivo uptake with respect to ex-vivo measurements, four metrics were estimated. To assess raw correlation between the measurements, Pearson's correlation coefficient was estimated. Second, intra-class correlation coefficients (ICCs) were estimated from linear mixed models. In this context, ICC is the estimated ratio of between-animal variance over the total variability observed (sum of between-animal and between-modality variance), a larger ICC (closer to one) represents smaller error between the *in vivo* and *ex vivo* measurements. To quantify how close the *in vivo* and *ex vivo* measurements were (on the units of observation), a repeatability coefficient was calculated. By construction, this value represents the value at which we would expect pairwise differences (*in vivo* measurement – *ex vivo* measurement) to fall under, with 95% probability. A smaller repeatability coefficient is desired. Finally, to determine if there was evidence of systemic bias between the in-vivo and ex-vivo measurements, paired t-tests were calculated. A smaller p-value provides evidence against the null hypothesis that the mean pairwise difference is zero.

To assess the variance in time-to-tumor-tripling between the dosage groups, a Cox proportional hazards model was fit to both the SCC2 and TU-138 data. Estimated hazard ratios and associated 95% confidence intervals can be seen in Figure 4. A likelihood ratio p-value was calculated from a linear trend test to assess if there was evidence of a linear decrease in the hazard ratio as dosage increased.

To model the clonogenic survival data, a linear-quadratic protracted radiation model was fit to the surviving fraction (SF) data using non-linear least-squares. The estimated predictive error was calculated using a second-order Taylor expansion (*34*).

**RESULTS**

### Recovery of $^{124}$I in microPET/CT

A PVC phantom was used to assess and validate the preclinical microPET/CT scanners ability to perform quantitative imaging of $^{124}$I in spherical target volumes of varying size (Figure 1). The phantom was filled with a solution of $^{124}$I at a concentration of 298 kBq/ml and scanned for 120 million coincidence events. PET images were then reconstructed with the same parameters used for mice studies. ROIs drawn on the corresponding CT image volume for each of the nine spheres were used to measure observed activity concentration in the coregistered and resampled PET image. Recovery coefficients (RC) were then calculated using Equation 1 and plotted against sphere diameter (Figure 1C). RC was fit to a sigmoidal function. The largest sphere at 25 mm in diameter exhibited the highest recovery at 0.823. Recovery of $^{124}$I began to drop appreciably in spheres at or below 9.8 mm in diameter.

### CLR 124 Targets Head and Neck Cancer Cells *In vivo*

Longitudinal PET/CT imaging studies over the course of 6 d demonstrated the capacity of CLR 124 to target HNC and established an *in vivo* model for CLR 131 tumor dosimetry (Figure 2). Maximum intensity projections of PET images in representative mice show selective uptake and retention of CLR 124 in human xenograft HNC models and limited uptake in normal tissues (Figure 2a). The biodistribution of CLR 124 in normal tissues (Supplemental Figure S1) indicated hepatobiliary clearance of the agent driven by elevated heart (blood pool) activity, as previously reported (*21,35,36*). ROI analysis of tumor uptake at 1, 24, 48, 72, and 144 h after injection showed peak uptake of CLR 124 was achieved by 48-72 h with durable retention through 144 h. The cohorts exhibiting highest uptake were SCC-22B (4.37 ± 0.81 %IA/g at 144 h) and UW-13 (4.23 ± 0.41 %IA/g at 144 h). However, uptake in the remaining PDX models (Figure 2c, Supplemental Figure S2b) was notably lower than that observed in cell line derived xenografts (Figure 2b, Supplemental Figure S2a), with UW-36 tumors showing little to no appreciable uptake of CLR 124 (max of 1.98 ± 0.23 %IA/g at 48 h). Accounting for PVC, the 144 h uptake for cell line derived xenograft and PDX models was 5.67 ± 1.62 %IA/g and 4.19 ± 1.60 %IA/g, respectively. The average tumor volume at the time of imaging was 381 ± 253 mm$^3$ with PDX tumors experiencing slower growth (80 ± 93 mm$^3$) over the course of 6 d than that observed in cell line derived xenografts (108 ± 149 mm$^3$).

*Ex vivo* biodistribution studies were performed in a subset of mice (SCC-6, SCC-22B, Tu-138, UW-1, UW-13, and UW-64) immediately following the final imaging time point to verify *in vivo* tumor uptake. Uptake measurements via *ex vivo* biodistribution confirmed relative trends observed *in vivo* with the highest uptake again observed in the SCC-22B (8.01 ± 1.55 %IA/g) and UW-13 (9.71 ± 1.06 %IA/g) cohorts.

**Accuracy of *In vivo* Uptake Improves with Partial Volume Correction**

For a subset of mice used in CLR 124 imaging and dosimetry studies (n = 24), tumor tissue was excised following the final imaging time point (144 h) and an additional *ex vivo* biodistribution measure of uptake was acquired. Partial volume correction using phantom measured RC was applied to *in vivo* uptake based on the mean segmented tumor volume observed over the course of the imaging study. Tumor volumes in mice ranged between 4.8 and 11.6 mm in diameter, assuming spherical geometry. The accuracy of *in vivo* uptake improved significantly (Figure 3A) following PVC with the absolute difference between *in vivo* and *ex vivo* measurements being reduced from 3.50 %IA/g to 0.99 %IA/g. While the paired t-test p-value (p<.0001) suggests evidence of a systematic bias between the initial *in vivo* and *ex vivo* measurements, following PVC, the p-value increased (p>0.05), indicating a lack of evidence for systematic bias once PVC is applied. The Intraclass correlation between *in vivo* and *ex vivo* uptake increased from 0.24 to 0.77 after PVC, and the repeatability coefficient decreased from 0.05 to 0.03 following PVC. The three metrics above provide evidence of increased reproducibility and repeatability of the PVC. The magnitude of *ex vivo* uptake was, on average, 48 ± 14% higher than that of PET-based measures (Figure 3B) with the largest difference in uptake (66.8% and 67.4%) observed in the smallest tumor volumes (74 and 99 mm$^3$, respectively). This volume dependent variability in *in vivo* uptake was reduced following PVC. Inadequate resection of one SCC-6 tumor resulted in artificially low *ex vivo* uptake and has been excluded from this analysis. Figures and analysis including this datapoint are shown in Supplemental Figure S3.

### Theranostic CLR 124/131 Tumor Dosimetry

Longitudinal PET/CT imaging studies of CLR 124 in mice (n = 2-6) was used to estimate tumor dosimetry for a single fraction of CLR 131 in each xenograft tumor model. Estimates of cumulative absorbed dose prescriptions (Gy/MBq) of CLR 131 RPT are shown in Table 1. Tumor models that demonstrated the highest uptake of CLR 124, SCC-22B and UW-13, also exhibited the highest CLR 131 prescription doses (1.31 ± 0.27 and 1.24 ± 0.08 Gy/MBq, respectively). Given the improved quantification of *in vivo* uptake with PVC, we further applied RC factors to the *in vivo* dosimetry results using Equation 3. Following PVC, CLR 131 prescription doses for tumors increased by 64-125% (mean, 87 ± 17%). Overall trends observed across tumor models remained consistent following PVC. On average, the corrected tumor dose for cell line derived xenograft models, 1.75 ± 0.44 Gy/MBq, was 24% higher than that observed in PDX models, 1.40 ± 0.46 Gy/MBq.

### Tumor Growth Inhibition Increases with CLR 131 Absorbed Dose

A comprehensive study was performed to probe the response of HNC tumors to escalating levels of CLR 131 radiation dose relative to that of conventional XRT. CLR 131 was prescribed at 2, 4, 6, and 8 Gy using non-PVC theranostic dosimetry results. The radiation dose delivered based on PVC dosimetry was 3.3, 6.6, 9.8, and 13.1 Gy for SCC-2 or 3.4, 7.0, 10.4, and 13.9 Gy for Tu-138, which maintains discrete and comparable dose levels. Figure 4 provides a summary of tumor dose response in SCC-2 and Tu-138. Overall, significant growth delay is observed in SCC-2 tumor models with mean tumor growth for the 8 Gy CLR 131 cohort following closely with 6 Gy XRT (Figure 4A). To assess the variance in growth across the treatment groups, a time-to-event analysis, where the event was defined as tumor-tripling, was performed via a Cox proportional hazards regression model. From this model, it was found that the 6 Gy XRT cohort exhibited the largest deviation from control in both SCC-2 and Tu-138 (Figure 4B&E). The response of each cohort varied relative to control over the course of observation, though this variation failed to reach statistical significance given the high level of within-group heterogeneity observed. Figure 4There is a slight linear relationship between radiation dose and TT-free survival for SCC-2 ($p < 0.02$) with respect to estimated hazard ratios (HR) (Figure 4C), but no statistically significant correlation observed for Tu-138 despite the data trending towards lower HR with increasing dose. For both models, the 6 Gy XRT cohort exhibited statistically significant response ($p < 0.005$). However, only the 8 Gy group for SCC-2 showed significant response to CLR 131 ($p < 0.02$). In all CLR 131 treatment groups, a HR of less than 1 was observed.

Clonogenic assays of SCC-2 and Tu-138 showed markedly different in vitro response to XRT with SCC-2 cells responding more severely than Tu-138 for the same radiation dose (Supplemental Figure 4). The α/β parameters derived from the linear-quadratic model was 1.99 Gy for SCC-2 and 2.58 Gy for Tu-138, both indicative of late responding tissues. In clinical terms, a lower α/β ratio indicates a higher biological effective dose for a given fractionation scheme with conventional XRT.

**DISCUSSION**

Reirradiation of recurrent and unresectable HNC has proven to be an effective, yet challenging, treatment approach in the clinic. Advances in conformal XRT have improved disease control and survival, but high risk of significant locoregional toxicity remains dose-limiting. In recent years there has been a growing interest in the use of RPT agents, with toxicity concerns oriented towards red marrow and clearance organs, in combination with conventional therapies (*37*). Used in tandem, RPT with agents such as CLR 131 can boost dose locoregionally and target systemic metastatic disease while lowering requisite dose from XRT and associated locoregional toxicity. However, despite HNC emerging early on as a promising target (*38*), the capacity for CLR1404 to broadly target HNC and deliver effective radiation dose has not been thoroughly studied. In this work, we employed the pairing of CLR 124 and CLR 131 to investigate the extent at which radioiodinated CLR1404 impinges upon HNC *in vivo* and probed the feasibility of theranostic dosimetry with incorporation of PVC to inform the strategic delivery of CLR 131 RPT.

PET/CT imaging of CLR 124 in flank tumor bearing mice showed selective uptake and retention in a broad spectrum of both cell line- and patient-derived xenograft HNC models. The biodistribution of CLR 124 in normal tissues, as reported in Supplemental Figure S1, demonstrated hepatobiliary clearance of CLR 124 and prolonged retention in blood aligns with observations from prior preclinical investigations of CLR 124 (*21,35,36*). Notably, PDX models exhibited 26% lower uptake than that of cell line derived models, on average, despite UW-13 demonstrating the second highest uptake of all tumor models. The source of this discrepancy is multifaceted. As posited by Cosper et al., the tumor stromal microenvironment and confluence of human and mouse tissue affects tumor vascularization and physiology in such a way that makes drug delivery challenging in these PDX models (*27*). Our observation of slower tumor growth in PDX models is indicative of this. However, robust uptake in UW-13 challenges this notion and suggests that affinity for CLR1404 may vary depending on individual cancer strain morphology independent of the nominal *in vivo* behavior of PDX models. These observations imply that pre-therapy confirmation of uptake would be beneficial in screening patients for CLR 131 RPT. It is also worth noting that the two HPV+ models, UW-1 and UW-36, exhibited the lowest uptake. Although far from conclusive, these results indicate that the impact of HPV status on CLR1404 targeting of HNC cancer should be further investigated.

The challenge of imaging $^{124}$I with quantitative accuracy has long been an issue in nuclear medicine and is becoming increasingly important to address with the resurgence of theranostics (*39,40*). In this work, considerable PVE's were observed in quantitative *in vivo* uptake measurements of CLR 124 such that the concentration of $^{124}$I activity was systematically underreported and that this bias grew increasingly large for small tumors. Employing a novel preclinical hot-sphere phantom, we found that the application of RC factors, as a function of measured spherical tumor volume, significantly improved the accuracy of *in vivo* quantification of CLR 124. Our results match up well with the observations of Knowles et al (*41*) who investigated $^{124}$I-A11 in prostate cancer and similarly assessed the accuracy of *in vivo* uptake measures with PVC using *ex vivo* biodistribution data acquired immediately following imaging. Notably, Knowles et al. utilized background subtraction prior to application of PVC using an ROI encompassing nonspecific tissue near the tumor. While this approach helps account for the assumption of a cold background that RC values are measured in, this tumor to background ratio changes with time after injection. This makes background subtraction more difficult to apply to cumulative measures such as absorbed dose, which is the primary focus of this work.

Given the improved quantification of CLR 124 uptake we observed with PVC, RC-based corrections were carried through to the CLR 124 PET-based theranostic tumor dosimetry for CLR 131. We found that when PVC was performed the radiation dose delivered to tumors per injected activity increased by 64-125%. Notably, the radiation dose delivered to tumors by CLR 131 RPT varied substantially between HNC models (0.89 to 2.25 Gy/MBq), emphasizing the importance of personalized tumor dosimetry CLR 131 RPT. However, we also found that tumor dose guided delivery of CLR 131 RPT achieves significantly different tumor response between models. It has largely been accepted that a Gy of XRT is not biologically equivalent to a Gy of RPT, but the work presented here further demonstrates that the tumoricidal capacity of a Gy of CLR 131 RPT can vary between HNC cell lines as well. We determined that for SCC-2 tumors, 13.1 Gy from CLR 131 will produce a similar response to 6 Gy of XRT, which is not true for Tu-138 tumors, where a Gy of CLR 131 achieves less relative effectiveness than XRT. A slightly higher α/β value for SCC-2 (1.99 Gy) compared to Tu-138 (2.58 Gy) indicates that a higher biologically effective dose delivered to SCC-2 may explain this variability. This provides further evidence that, when it comes to dosimetry, radiobiological factors need to be accounted for in order to fully understand and leverage the prognostic power of radiation dosimetry (*42*).

There are a few notable limitations of this work. While the application of RC-based PVC for *in vivo* CLR 124 uptake was validated with direct *ex vivo* biodistribution measurements, the correction holds only for region-level metrics such as mean absorbed dose. The approach assumes uniform uptake within the target volume. This is often not the case in larger tumors with spatially heterogeneous or concentrated manifestation of hypoxia. Furthermore, using region-based PVC to arrive at more accurate mean dose metrics curtails much of the valuable information that can be gleaned from the personalized voxel-level dosimetry approach employed in this work. Given the importance of voxel-level dosimetry in clinically relevant radiobiological factors, the translational potential of this PVC approach is limited to small lesions. Another limitation of this work lies in the variability in tumor size in preclinical imaging and therapy studies. Our incorporation of volume-based PVC reduced the impact of this variability on uptake and dosimetry analysis. For therapy studies, however, a considerable range in initial tumor volumes (72 to 1360 mm$^3$) introduced uncertainty to our response analysis.

In our statistical analysis, this manifested as a high degree of within-group heterogeneity in the longitudinal growth models, which proved challenging for the many internal optimization algorithms used in the fitting of hierarchical statistical models. However, this variability in tumor size is often observed in preclinical studies and efforts to reduce this would require preparation of a prohibitively large number of mice per experiment.

**CONCLUSIONS**

The work presented herein establishes that CLR 124/131 demonstrates tumor specific uptake and durable retention in a broad spectrum of HNC xenograft models. Partial volume effects significantly impact quantitative PET imaging of CLR 124 in tumors and the methodology validated here for incorporating PVC in dosimetry should be considered in future theranostic applications of CLR 124/131 and other $^{124}$I-based agents. The tumor-dosimetry driven response studies presented here suggest that CLR 131 has the capacity to deliver increasingly tumoricidal doses to HNC tumors and that careful consideration of radiobiological factors should be made when considered in the context of conventional XRT. This work further supports and advances efforts to clinically translate combination RPT+XRT treatments for patients with HNC.

**DISCLAIMER**

Research reported in this publication was supported by the National Cancer Institute (NCI) of the National Institutes of Health (NIH) under Award Number T32CA009206 and the Specialized Program of Research Excellence (SPORE) program, through the NIH National Institute for Dental and Craniofacial Research and NCI grant P50DE026787. Support was also provided by grants U01CA233102-01 and P01CA250972-01. The content is solely the responsibility of the authors and does not necessarily represent the official views of the National Institutes of Health.

Jamey Weichert was cofounder of Cellectar Biosciences and Justin Jeffery is a cofounder of Phantech.

**ACKNOWLEDGEMENTS**

We would like to acknowledge the Caner Center Support Grant NCI P30 CA014520 and thank the University of Wisconsin Small Animal Imaging & Radiotherapy Facility and The Ride (Madison, WI) student scholar award for supporting this work.

**KEY POINTS**

**Question:**

Does CLR1404 have the potential to target head and neck cancer *in vivo* and can CLR 124-based theranostic dosimetry be used for dosimetry guided RPT with CLR 131?

**Pertinent Findings:**

CLR 124 shows selective uptake and retention in a broad spectrum of cell line and patient derived xenograft models of head and neck cancer. The theranostic combination of CLR 124/131 allows for dosimetry guided RPT, but efforts should be made to account for radiobiological factors and $^{124}$I partial volume effects.

**Implications for Patient Care:**

CLR 131 shows promise for further clinical translation against head and neck cancer with significant potential for use in combination with conventional XRT to help reduce locoregional toxicity.

**FIGURES AND TABLES**

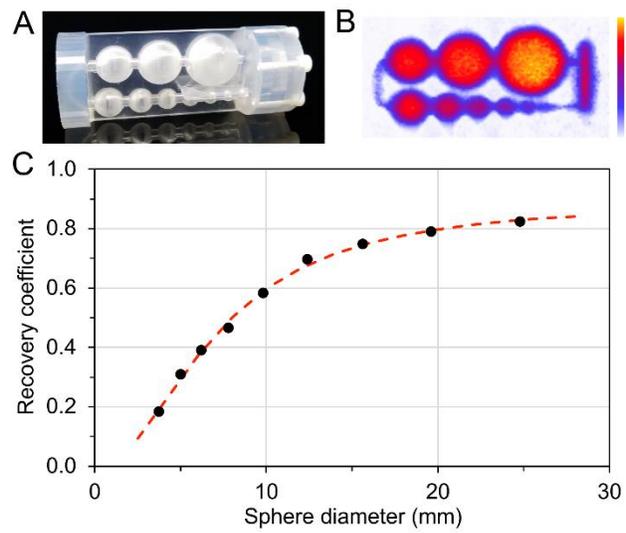

Figure 1: Partial volume correction phantom analysis for $^{124}$I in microPET/CT imaging. (**A**) Photograph of the partial volume correction phantom used in this work showing a series of 9 hollow spheres ranging from 28 to 8000 mm$^3$. (**B**) Normalized maximum intensity projection of the PET imaged phantom. (**C**) Recovery of $^{124}$I activity vs sphere diameter as measured by ROI analysis using contours drawn on the corresponding CT image.

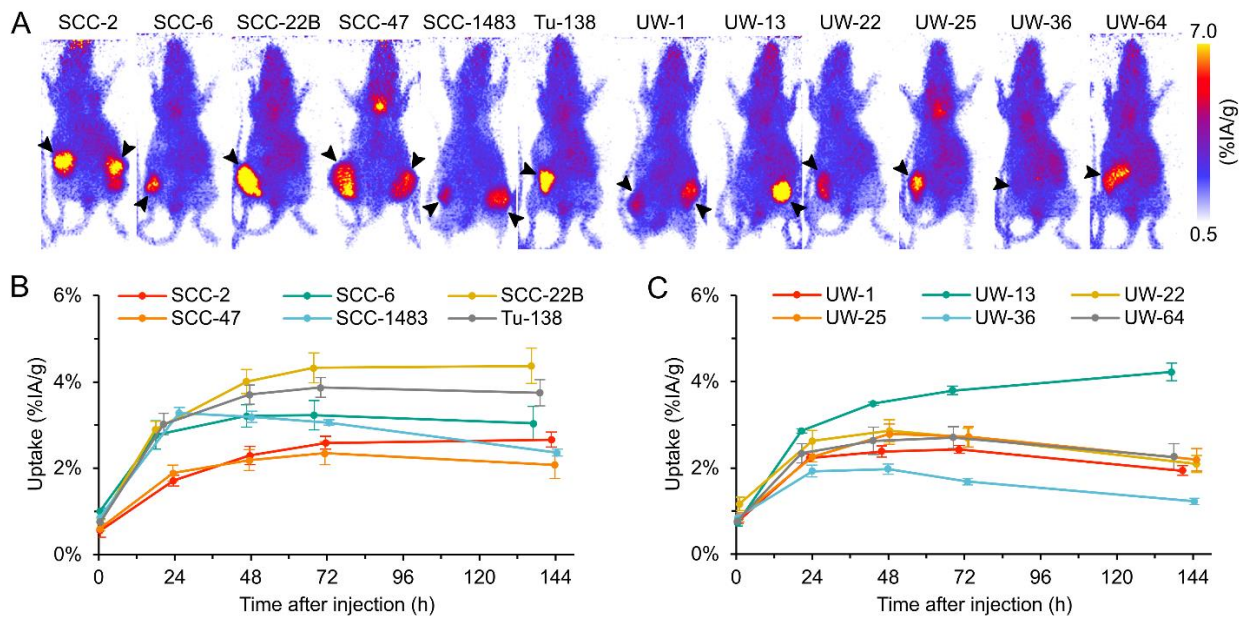

Figure 2: PET/CT imaging and ROI analysis of flank tumor bearing mice following intravenous administration of CLR 124. (**A**) PET maximum intensity projections of representative mice in each cohort at 144 h after injection. Black arrows indicate tumor location. *In vivo* time activity curves of CLR 124 uptake in (**B**) cell line and (**C**) patient derived xenograft tumor models. Data presented is in terms of mean percent injected activity per gram (%IA/g) with standard error bars shown (n = 2-6).

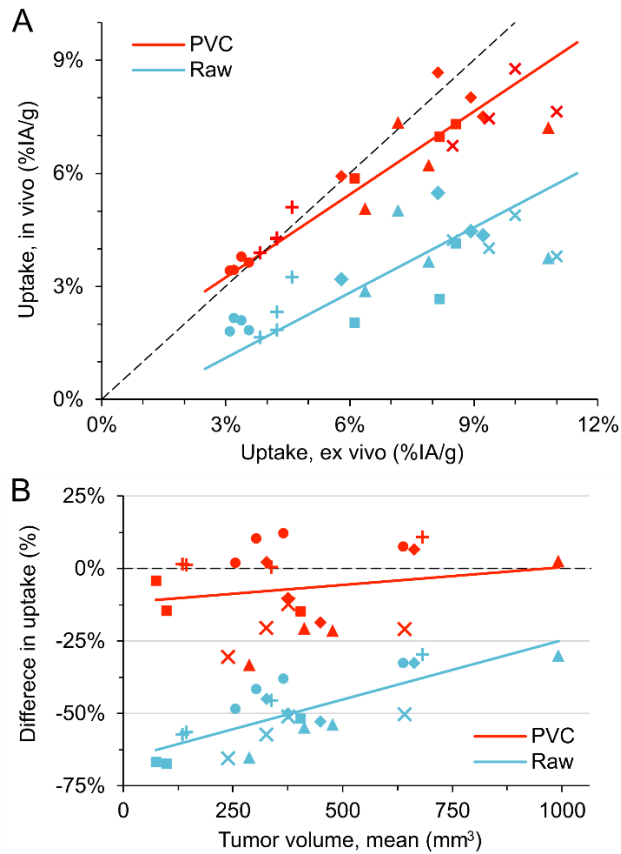

Figure 3: Impact of PVC on *in vivo* uptake measures relative to *ex vivo*. (A) Tumor uptake of CLR 124 approximately 144 h after injection as measured *in vivo* and *ex vivo*. *In vivo* uptake is shown both without (blue) and with (red) PVC. The black-dashed line represents 1:1 measure of *in vivo* and *ex vivo* uptake. (B) The difference between *in vivo* and *ex vivo* uptake for raw and PVC relative to average tumor volume. SCC-6 (■), SCC-22B (◆), Tu-138 (▲), UW-1 (●), UW-13 (✕), and UW-64 (+) tumor models are shown.

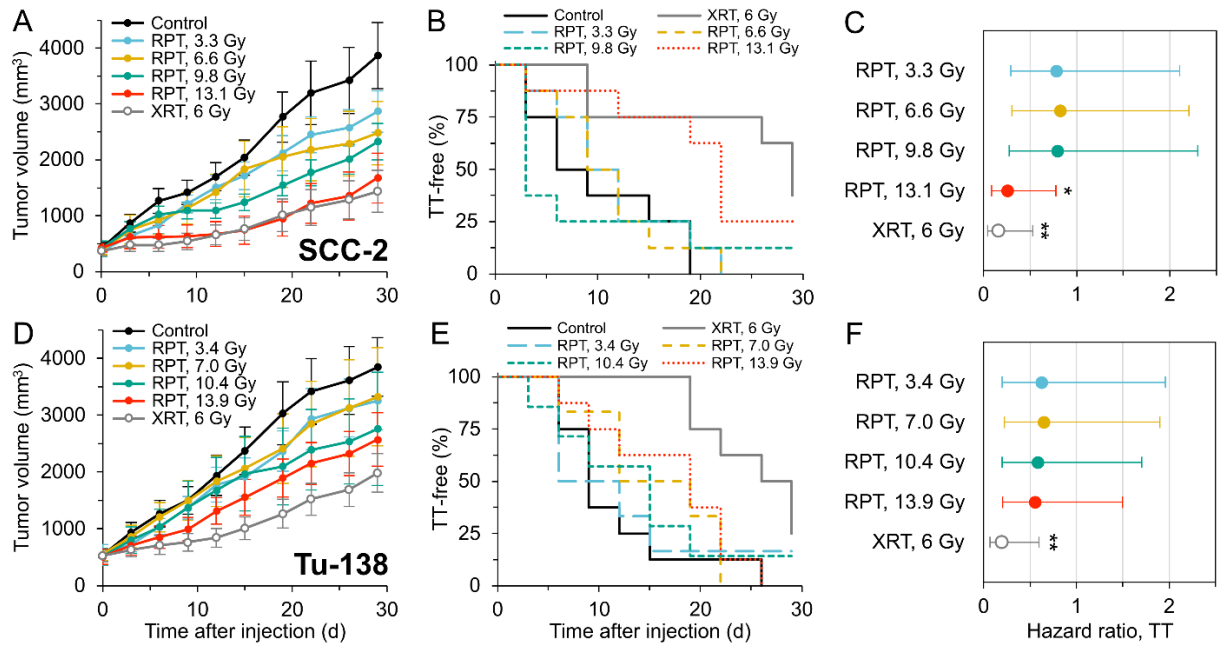

Figure 4: Dose escalation and tumor response to CLR 131 in SCC-2 (A-C) and Tu-138 (D-F) tumor bearing mice. (A,D) Tumor growth in mice (mean ± SE, n = 6-8) treated with CLR 131 at four tumor-dose levels based on prior theranostic dosimetry using CLR 124. Untreated (Control) and 6 Gy XRT cohort shown for reference. (B,E) Time-to-event analysis shown using tumor-tripling (TT) as the endpoint. (C,F) Overall hazard-ratio for each treatment cohort with error bars representing the 95% CI (*$p < 0.02$, **$p < 0.005$).

**TABLES**

Table 1: Subject-specific tumor dosimetry for CLR 131 in tumor bearing mice

| Xenograft | Cumulative absorbed dose Rx (Gy/MBq) | |
| --- | --- | --- |
|  | raw | with PVC |
| SCC-2 | 0.78 ± 0.12 | 1.28 ± 0.15 |
| SCC-6 | 0.97 ± 0.24 | 2.19 ± 0.22 |
| SCC-22B | 1.31 ± 0.27 | 2.25 ± 0.33 |
| SCC-47 | 0.65 ± 0.18 | 1.27 ± 0.25 |
| SCC-1483 | 0.83 ± 0.07 | 1.55 ± 0.27 |
| Tu-138 | 1.16 ± 0.25 | 1.94 ± 0.47 |
| UW-1 | 0.64 ± 0.08 | 1.16 ± 0.10 |
| UW-13 | 1.24 ± 0.08 | 2.24 ± 0.20 |
| UW-22 | 0.70 ± 0.15 | 1.32 ± 0.21 |
| UW-25 | 0.70 ± 0.17 | 1.36 ± 0.23 |
| UW-36 | 0.45 ± 0.05 | 0.89 ± 0.15 |
| UW-64 | 0.75 ± 0.23 | 1.46 ± 0.15 |